\documentclass[a4paper]{spie}  
\usepackage[]{graphicx}
\usepackage{amssymb,amsmath,psfrag,url}
\usepackage{verbatim}

\newcommand{\Bolivarallee}{Boliva\hspace{-0.1mm}r\hspace{0.15mm}a\hspace{-0.1mm}llee}
\newcommand{\Abbestrasse}{Abbe\hspace{0.25mm}s\hspace{-0.1mm}tra{\ss}e}
\newcommand{\Takustrasse}{Taku\hspace{0.25mm}s\hspace{-0.1mm}tra{\ss}e}

\title{Investigation of 3D Patterns on EUV Masks by Means of Scatterometry and Comparison to Numerical Simulations}

\author{
Sven~Burger,\supit{\,ab}
Lin~Zschiedrich,\supit{\,b}
Jan~Pomplun,\supit{\,b}
Frank~Schmidt,\supit{\,ab}
Akiko~Kato,\supit{\,c}
Christian~Laubis,\supit{\,c}
Frank~Scholze\supit{\,c}
\skiplinehalf
\supit{a}
Zuse Institute Berlin\,(ZIB),
\Takustrasse~7,
D\,--\,14\,195 Berlin,
Germany
\smallskip\\
\supit{b}
JCMwave GmbH,
\Bolivarallee~22, 
D\,--\,14\,050 Berlin,
Germany
\smallskip\\
\supit{c}
Physikalisch-Technische Bundesanstalt\,(PTB), 
\Abbestrasse~2\hspace{0.5mm}--\,12, 
D\,--\,10\,587 Berlin, 
Germany
}
\authorinfo{
Corresponding author: S. Burger\\
URL: http://www.zib.de\\
URL: http://www.jcmwave.com\\
Email: burger@zib.de
}

\begin{document}
\maketitle
\noindent
This paper will be published in Proc.~SPIE Vol. {\bf 8166}
(2011) 81661Q 
({\it Photomask Technology 2011; W.~Maurer, F.E.~Abboud, Editors}, 
DOI: 10.1117/12.896839), 
and is made available 
as an electronic preprint with permission of SPIE. 
One print or electronic copy may be made for personal use only. 
Systematic or multiple reproduction, distribution to multiple 
locations via electronic or other means, duplication of any 
material in this paper for a fee or for commercial purposes, 
or modification of the content of the paper are prohibited.

\begin{abstract}
EUV scatterometry is performed on 3D patterns on EUV lithography masks. 
Numerical simulations of the experimental setup are performed using a rigorous Maxwell solver. 
Mask geometry is determined by minimizing the difference between experimental results and numerical results for 
varied geometrical input parameters for the simulations. 
\end{abstract}

\keywords{EUV scatterometry, optical metrology, 3D rigorous electromagnetic field simulations, computational lithography, finite-element methods}

\section{Introduction}
Extreme ultraviolet (EUV) lithography at a wavelength of about 13\,nm 
is expected to replace deep ultraviolet (DUV) lithography for fabrication of
integrated circuits with minimum feature sizes (critical dimension, $CD$) 
as small as 22\,nm or below. 
With decreasing feature sizes tolerance budgets of mask pattern 
dimensions get tighter. 
Metrology of such structures has to be performed for characterization and process control. 

In previous works it has been shown that EUV scatterometry is a fast and 
robust method for characterizing masks with 1D-periodic patterns 
(line masks)~\cite{Pomplun2006bacus,Gross2009euv1,Scholze2008a,Pomplun2008pmj}. 
In this contribution we extend our research to the characterization 
of 2D-periodic patterns (array of contact holes). 

In our method we compare the distribution of EUV 
light scattered off the sample to results from numerical simulations 
of the setup with different sets of geometrical parameters. 
Mask geometry is then determined by optimizing geometrical input 
parameters of the numerical simulations such that differences between 
measurement and simulation are minimized. 

This paper is structured as follows: 
The experimental setup of the EUV reflectometer at PTB is presented in Section~\ref{section_experimental}.
The used model of the investigated mask and the numerical method are described in Section~\ref{section_setup_mask}. 
Section~\ref{section_results} shows results on 3D geometry reconstruction. 

\section{Experimental EUV scatterometry results}
\label{section_experimental}
The data presented here were measured at the EUV reflectometry facility of PTB~\cite{Scholze2006a,Klein2006a} 
in its laboratory at the storage ring BESSY\,II. 
The PTB's soft X-ray radiometry beamline~\cite{Scholze2001a} uses a plane grating monochromator 
which covers the spectral range from 0.7\,nm to 35\,nm and which was particularly 
designed to provide highly collimated radiation. 
For this purpose it uses a long focal length of 8\,m in the monochromator, and 
the focusing in the non-dispersive direction is provided by a collecting 
pre-mirror with a focal length of 17\,m. 
We achieve a collimation of the radiation in the experimental station to 
better than 200\,$\mu$rad and the scatter halo of the beam can be suppressed 
to below $10^{-5}$ relative intensity at an angle of only 1.7\,mrad to the central beam. 

\begin{figure}[t]
\begin{center}
\begin{minipage}[c]{.36\textwidth}
  \includegraphics[width=.97\textwidth]{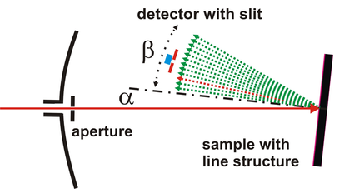}
 \end{minipage}
\begin{minipage}[c]{.62\textwidth}
  \includegraphics[width=.97\textwidth]{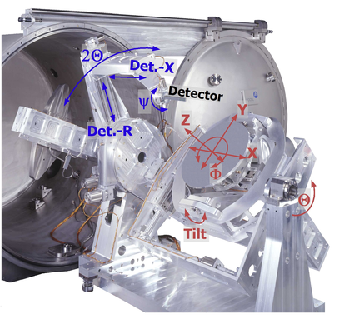}
 \end{minipage}
  \caption{
{\it Left:} Scheme of scatterometry measurements: 
The detector angle $\beta$ is scanned at a fixed incidence angle~$\alpha$.
{\it Right:} Photograph of the experimental setup. 
}
\label{figure_setup_experiment}
\end{center}
\end{figure}

The measurement scheme for angular resolved scatterometry of periodic structures is presented 
in Figure~\ref{figure_setup_experiment} ({\it left}). 
Usually, the sample is set at a fixed angle to the incoming radiation and the detector is moved. 
 Figure~\ref{figure_setup_experiment} ({\it right}) shows a photograph of the experimental setup: 
The probed photomask is mounted at the center of the (red marked) $x$-$y$-$z$-coordinate system.
It is rotated such 
that the periodicity vectors of the periodic patterns on the mask are in $x$- and $y$-directions. 
The EUV probe beam is incident in the $y$-$z$-plane. The detector can be freely moved to capture reflected 
$x$- and $y$-diffraction orders.

\begin{figure}[t]
\begin{center}
\begin{minipage}[c]{.45\textwidth}
  \psfrag{X}{\sffamily $N_\textrm{x}$}
  \psfrag{Y}{\sffamily $N_\textrm{y}$}
  \includegraphics[width=.97\textwidth]{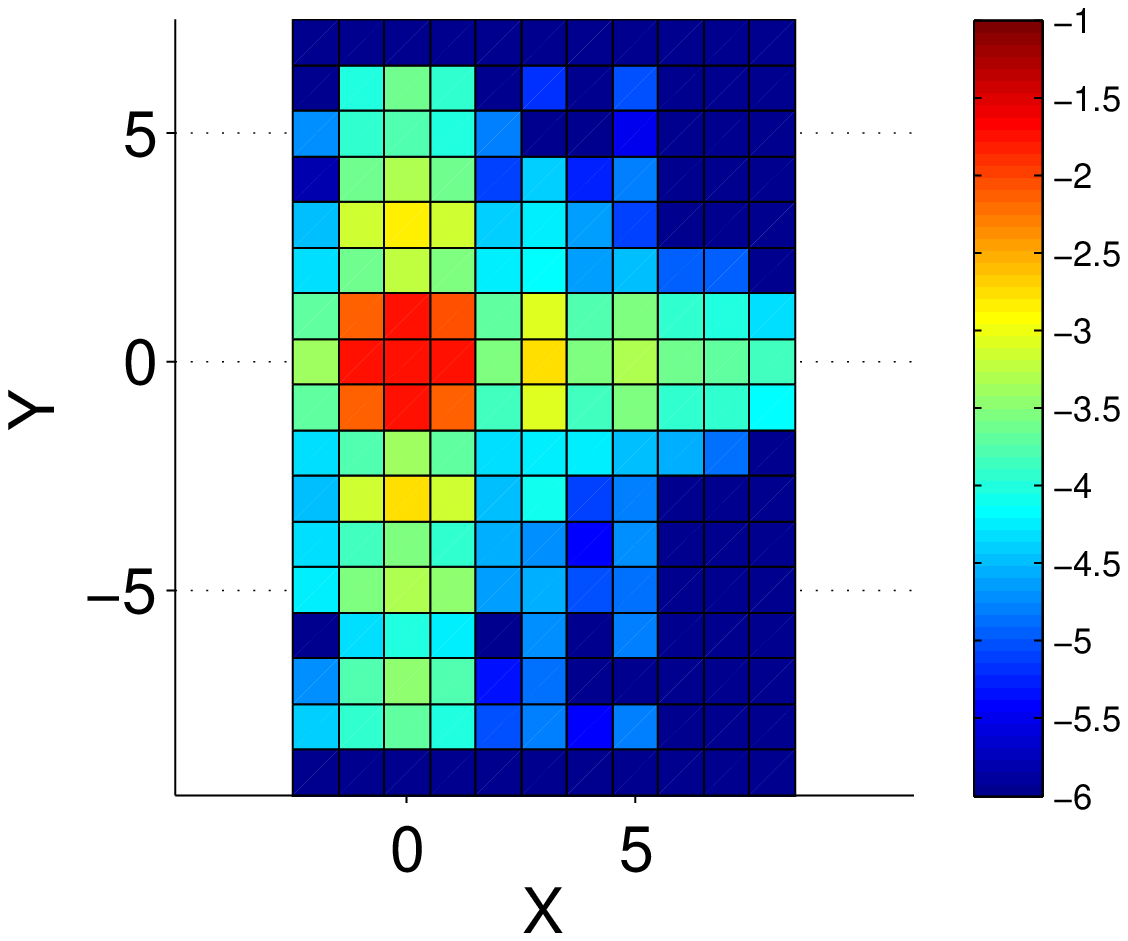}
 \end{minipage}
  \hfill
\begin{minipage}[c]{.52\textwidth}
  \psfrag{Diffraction orders}{\sffamily diffraction orders}
  \psfrag{Log(Intensities)}{\sffamily $\log_{10}(I)$}
  \includegraphics[width=.97\textwidth]{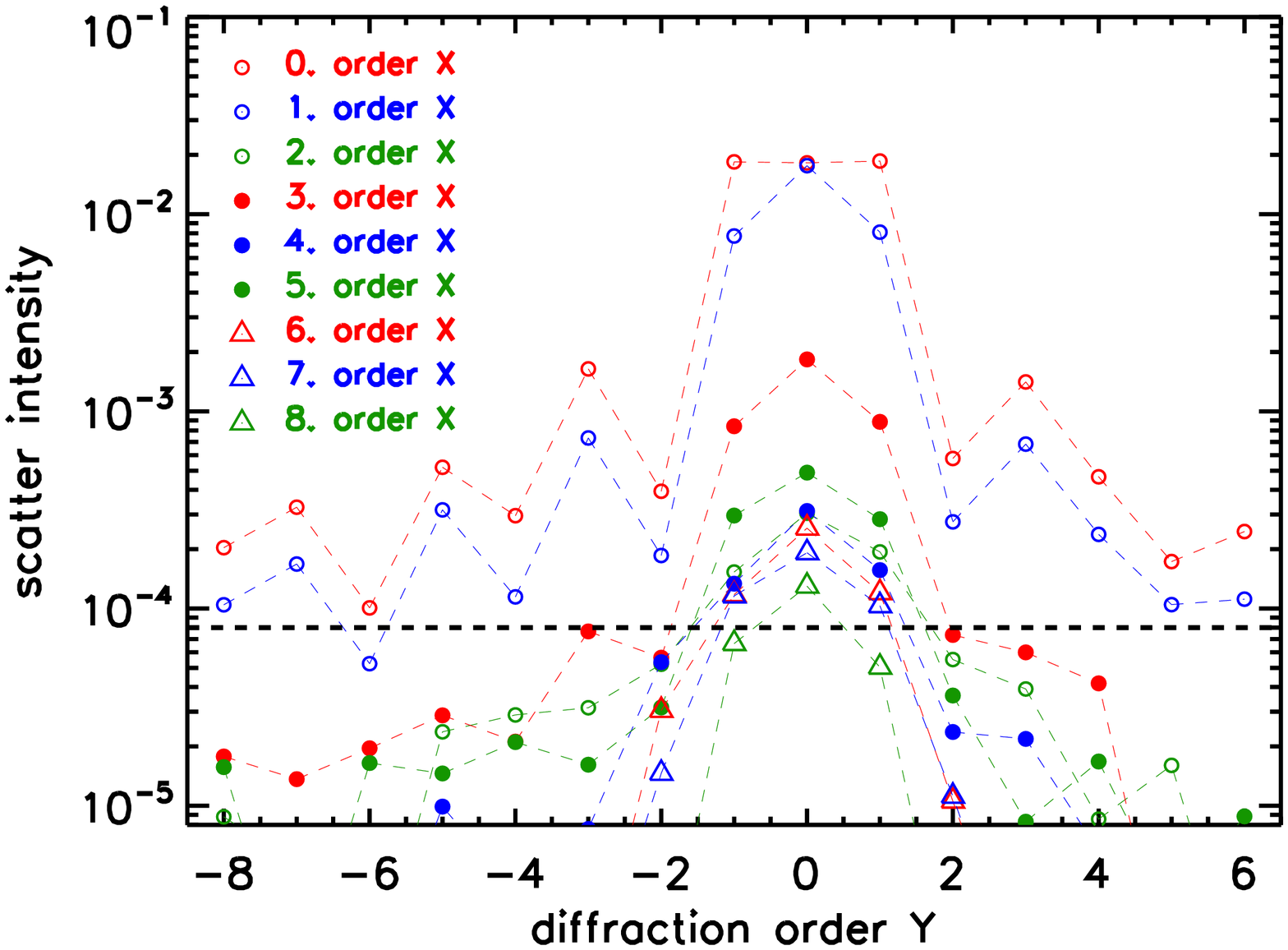}
 \end{minipage}
  \caption{
Scatterogram of an EUV mask with a contact hole array. 
{\it Left}: Measured diffraction intensities ($log_{10}(I^\textrm{exp}_{N_\textrm{x}, N_\textrm{y}})$) as function of diffraction index in $x$- and $y$-direction, 
$N_\textrm{x}$, $N_\textrm{y}$ in a pseudo-color representation.
{\it Right}: Same data as a 2D plot on a logarithmic scale.  
The dashed line indicates the threshold of $8\cdot 10^{-5}$ used in the evaluation.
}
\label{figure_experiment_CH}
\end{center}
\end{figure}

The presented experimental data are obtained from periodically structured 
areas on an EUV photomask. 
The structure for the measurements was a 
1.4\,mm by 1.4\,mm large field with a 2D structure 
of quadratic contact holes with nominal 300\,nm CD at 600\,nm pitch in $x$- and $y$-direction. 
 The measurement method used was angular resolved scatterometry. 
The scan range of the detector angle $\beta$ was from $1^\circ$ to $25^\circ$, 
see Figure~\ref{figure_setup_experiment}. 
The $-8^{\textrm th}$ to $+6^{\textrm th}$ diffraction orders were measured in $y$-direction 
with respect to the sample using the movement of the detector in the plane 
of specular reflection. 
For the measurement of diffraction in the out-of-plane direction, 
the perpendicular movement of the detector was used~\cite{Scholze2006a}.
Thus we also measured from the $-2^{\textrm nd}$ to $+8^{\textrm th}$ order in $x$-direction.
 As the EUV radiation is resonant with the underlying multilayer reflective coating, 
the measured diffraction intensities strongly depend on the actual wavelength. 
We measured at centre wavelength of the 
resonance peak of 13.54\,nm, see Figure~\ref{figure_multilayer_spectrum}. 
Figure~\ref{figure_experiment_CH} shows a scatterogram for a measurement of the 
contact hole array in two different graphical representations.

\section{Mask model and simulation method}
\label{section_setup_mask}
\subsection{Multilayer model}

\begin{figure}[t]
\begin{center}
  \includegraphics[width=.4\textwidth]{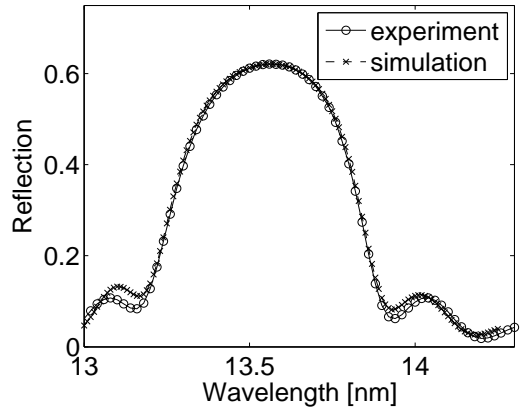}
  \caption{
Reflection from the blank multilayer mirror (without buffer, absorber and ARC):
Experimental results (o) and numerical results (x) for multilayer parameters as defined in 
Table~\ref{table_parameters}.
}
\label{figure_multilayer_spectrum}
\end{center}
\end{figure}

For characterizing the multilayer stack we have performed measurements of reflectivity off the 
blank multilayer mirror as a 
function of incident wavelength in a spectral range from 13\,nm to 14.3\,nm. 
The multilayer stack consists of 40 $\lambda/2$-multi-layers of molybdenum (Mo) and silicon (Si). 
It further contains a silicon capping layer, finished with a thin oxide layer, and it is 
placed on a SiO$_2$ substrate. 
To model the $\lambda/2$-multi-layers we have chosen a four-layer stack consisting of 
a Mo-layer, a MoSi$_2$-layer (interlayer), a Si-layer, and a second MoSi$_2$-layer. 
A schematic is shown in Figure~\ref{schematics_euv} ({\it left}). 
We have fitted the layer thicknesses of the blank mirror using the IMD multilayer programme~\cite{Windt1998a}, 
using also the therein provided material definitions. 
The resulting reflectivity spectrum is display in Figure~\ref{figure_multilayer_spectrum} 
together with measured data. 
The obtained layer thicknesses are shown in Table~\ref{table_parameters}.
These values are used in the subsequent 3D simulations.
For the thicknesses of the absorber stack (buffer, absorber and anti-reflection, ARC, layers) 
we have chosen the nominal values of the EUV mask (see Table~\ref{table_parameters}). 

\begin{table}[h]
\begin{center}
\begin{tabular}{|l|l|l|l|l|}
\hline
material & height & $n$ & $k$ & \\
\hline
vacuum & $\inf$ & 1 & 0 & \\
ARC & $h_{\textrm{ARC}} = 12\,$nm & 0.9474 & 0.0316 & $R_{\textrm{z}}=5\,$nm\\
absorber & $h_{\textrm{a}} = 55\,$nm & 0.9255 & 0.0439 &  \\
buffer & $h_{\textrm{b}} = 10\,$nm  & 0.9735 & 0.0131 & \\
oxide & $h_{\textrm{ox}} = 1.35\,$nm & 0.9735 & 0.0131 & \\
capping & $h_{\textrm{cap}} = 8.19\,$nm & 1.0097 & 0.0013 & \\  
multi-layer (interlayer) & $h_{\textrm{i}} = 1.48\,$nm & 0.9681 & 0.0044 &  40\,layers\\
multi-layer (Mo) & $h_{\textrm{mo}} = 2.99\,$nm & 0.9206 & 0.0065 & 40\,layers \\
multi-layer (interlayer) & $h_{\textrm{i}} = 1.48\,$nm & 0.9681 & 0.0044 &  40\,layers\\
multi-layer (Si) & $h_{\textrm{si}} = 1.18\,$nm & 1.0097 & 0.0013 &  40\,layers\\
substrate & $\inf$ & 0.9735 &  0.0131 & \\
\hline
\hline
lateral dimensions & $CD_{\textrm{bottom,x}}$ & \multicolumn{3}{l|}{300\,nm}\\
(nominal)          & $CD_{\textrm{bottom,y}}$ & \multicolumn{3}{l|}{300\,nm}\\
                   & $p_{\textrm{x}}$ & \multicolumn{3}{l|}{600\,nm}\\
                   & $p_{\textrm{y}}$ & \multicolumn{3}{l|}{600\,nm}\\
                   & $R_{\textrm{xy}}$ & \multicolumn{3}{l|}{0\,nm}\\
                   & sidewall angle $\alpha$ & \multicolumn{3}{l|}{90\,deg}\\
\hline
\hline
illumination & angle of incidence $\alpha_{\textrm{in}}$ & \multicolumn{3}{l|}{6\,deg}\\
                   & wavelength $\lambda_0$ & \multicolumn{3}{l|}{13.54\,nm}\\
                   & polarization & \multicolumn{3}{l|}{$E=(E_x, 0, 0)$}\\
\hline
\end{tabular}
\caption{Parameter setting for the EUV
mask simulations (compare Fig.~\ref{schematics_euv}):
Mask geometry parameters (layer heights $h_x$, sidewall angle $\alpha$, corner rounding 
radius $R_{\textrm{z}}$), material parameters 
(real and imaginary parts of the refractive index, $n$ and $k$, at a wavelength 
of 13.54\,nm), 
illumination parameters (in-plane angle of incidence, vacuum wavelength, plane-wave electric field polarization).
}
\label{table_parameters}
\end{center}
\end{table}


\subsection{Lateral layout: array of contact holes}

\begin{figure}[t]
\begin{center}
\psfrag{R}{\sffamily $R_\textrm{z}$}
\psfrag{alpha}{\sffamily $\alpha$}
\psfrag{harc}{\sffamily $h_{\textrm{ARC}}$}
\psfrag{ha}{\sffamily $h_\textrm{a}$}
\psfrag{hb}{\sffamily $h_\textrm{b}$}
\psfrag{hox}{\sffamily $h_{\textrm{ox}}$}
\psfrag{hcap}{\sffamily $h_{\textrm{cap}}$}
\psfrag{hmo}{\sffamily $h_{\textrm{mo}}$}
\psfrag{hsi}{\sffamily $h_{\textrm{si}}$}
\psfrag{hi}{\sffamily $h_{\textrm{i}}$}
\psfrag{ha}{\sffamily $h_\textrm{a}$}
\psfrag{arc}{\sffamily ARC}
\psfrag{absorber}{\sffamily absorber}
\psfrag{buffer}{\sffamily buffer}
\psfrag{oxide}{\sffamily oxide}
\psfrag{capping}{\sffamily capping}
\psfrag{multilayer}{\sffamily multi-layer (40 $\times$ 4)}
\psfrag{substrate}{\sffamily substrate}
\psfrag{px}{\sffamily $p_\textrm{x}$}
\psfrag{cd}{\sffamily $CD_{\textrm{bottom}}$}
\psfrag{Rxy}{\sffamily $R_{\textrm{xy}}$}
\psfrag{CDx}{\sffamily $CD_{\textrm{x}}$}
\psfrag{CDy}{\sffamily $CD_{\textrm{y}}$}
\psfrag{px}{\sffamily $p_{\textrm{x}}$}
\psfrag{py}{\sffamily $p_{\textrm{y}}$}
  \includegraphics[width=.4\textwidth]{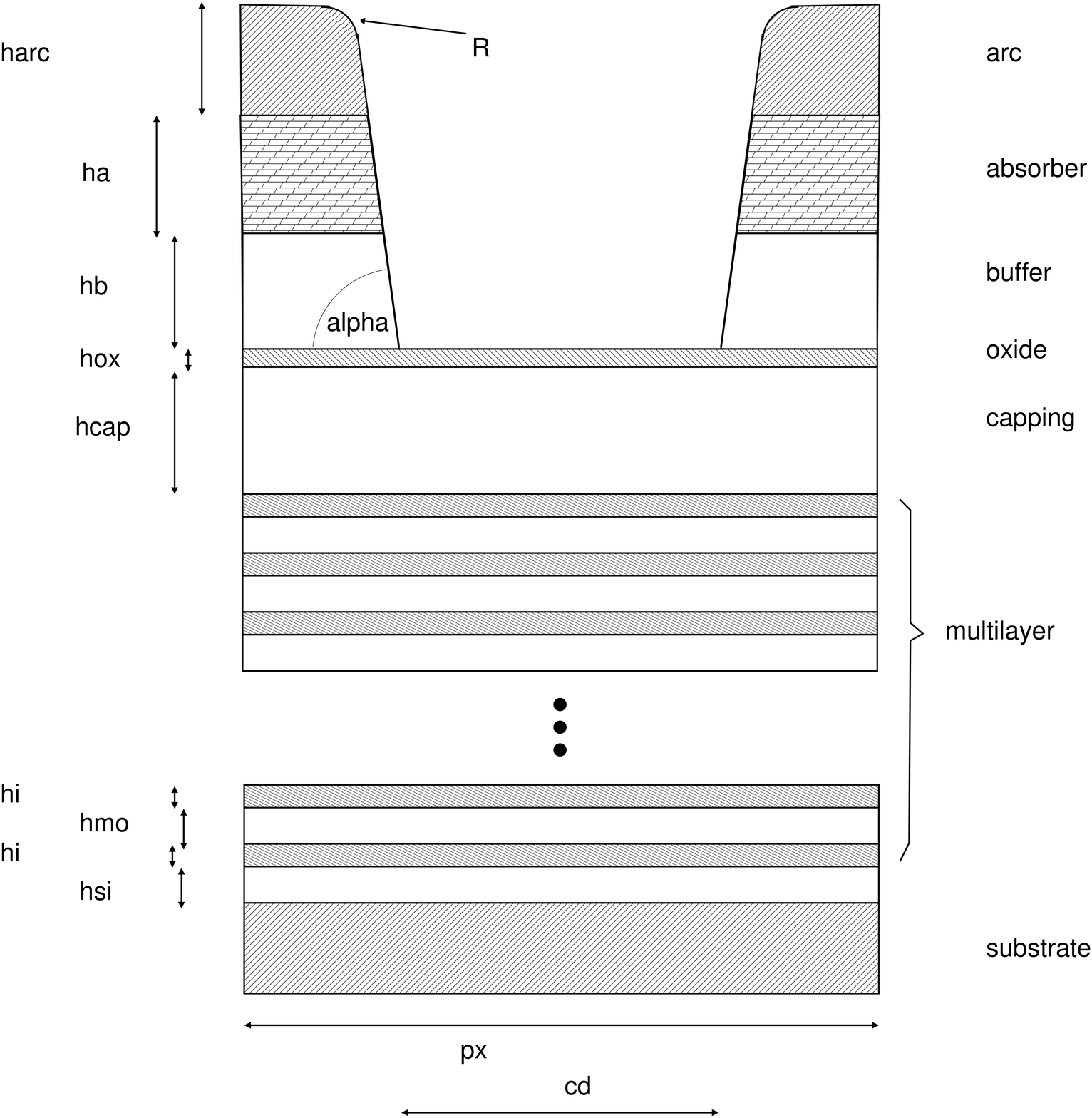}
  \hfill
  \includegraphics[width=.4\textwidth]{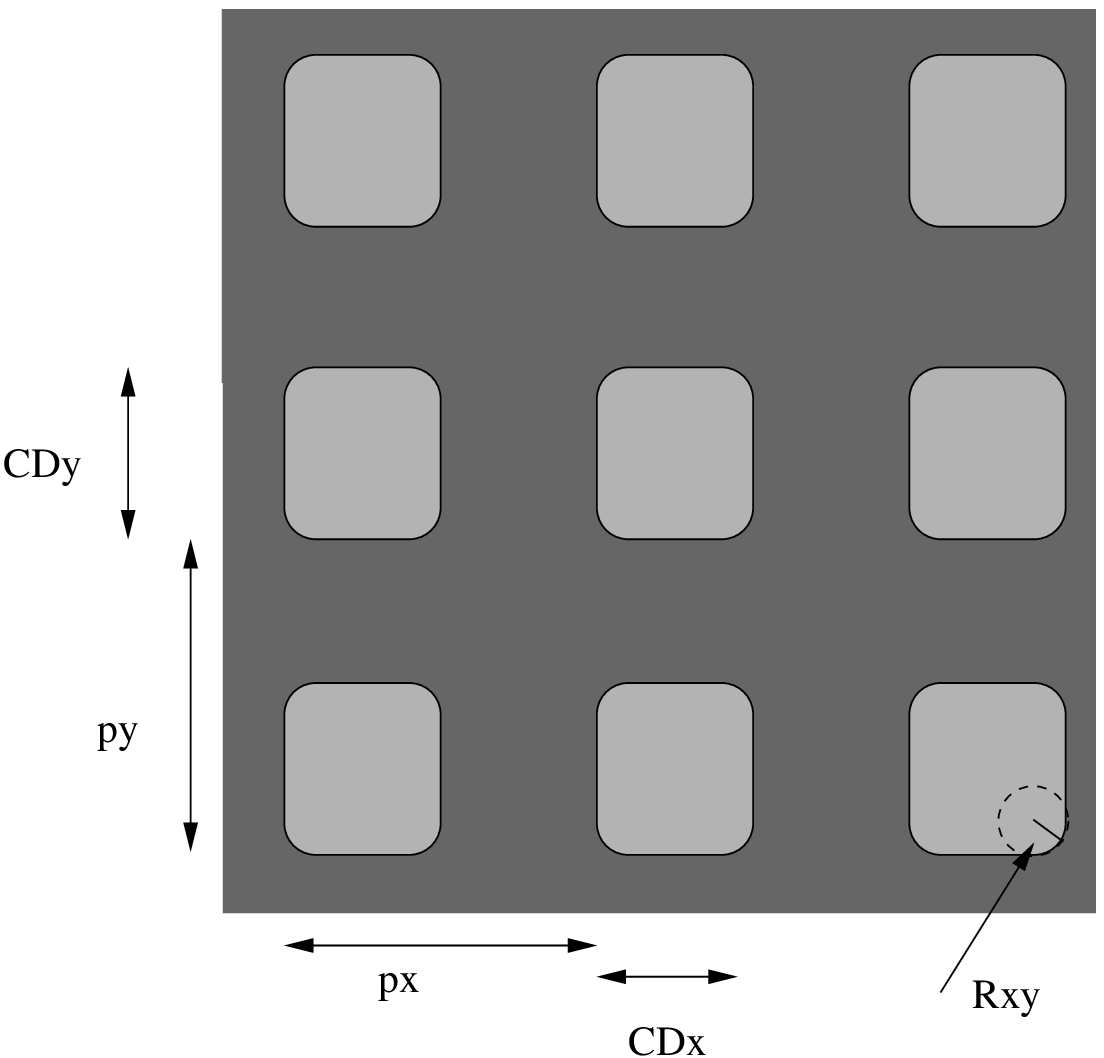}
  \caption{
{\it Left}: 
Schematic of a 2D cross-section through the 3D setup. 
{\it Right}: Top view ($x$-$y$-plane), schematic of the lateral geometry. 
For parameter settings, compare Table~\ref{table_parameters}.
}
\label{schematics_euv}
\end{center}
\end{figure}

The investigated mask holds several different patterns. 
Here we concentrate on analysis of a 
single of these patterns: an array of contact holes,  
Figure~\ref{schematics_euv} ({\it right}) shows a schematic of the lateral layout. 
A pattern with a pitch of $p_\textrm{x} = p_\textrm{y} = 600\,$nm is investigated. 
Nominal values for the (bottom) CD's are $CD_\textrm{x} = CD_\textrm{y} = 300\,$nm. 
We assume a sidewall angle of the absorber stack, $\alpha$, and a constant 
(vertical) corner rounding of the stack, $R_\textrm{z}$. 
Free parameters in our investigation are the contact hole critical dimensions $CD_\textrm{x}$, $CD_\textrm{y}$, the sidewall angle $\alpha$, 
and the lateral corner rounding radius, $R_{\textrm{xy}}$.

\begin{figure}[t]
\begin{center}
  \includegraphics[width=.4\textwidth]{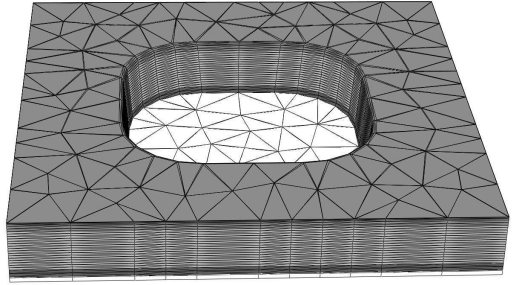}
  \caption{
Mesh of the structured absorber geometry. The discretizations of 
the adjacent vacuum region and of the multi-layer stack are not displayed. 
}
\label{figure_grid}
\end{center}
\end{figure}

\subsection{Simulation setup}
For rigorous simulations of the scattered EUV light field we use the 
finite-element (FEM) Maxwell solver JCMsuite.
This solver incorporates higher-order edge-elements, self-adaptive meshing, 
and fast solution algorithms for solving time-harmonic Maxwell's equations. 
Previously the solver has been used in scatterometric investigations   
of EUV line masks (1D-periodic patterns)~\cite{Scholze2007a,Scholze2008a}.
Recently we have reported on rigorous electromagnetic field simulations of 2D-periodic 
arrays of absorber structures on EUV masks~\cite{Burger2011eom1}.  
This report contained a convergence study which demonstrates that highly  accurate, rigorous 
results can be attained even for the relatively large 3D computational domains which are typically present 
in 3D EUV setups. 

Briefly, the simulations are performed as follows:
a scripting language (Matlab) automatically iterates the input parameter sets 
($CD_\textrm{x}$, $CD_\textrm{y}$, $\alpha$, $R_{\textrm{xy}}$). 
For each set, a prismatoidal 3D mesh is created automatically by the built-in mesh generator. 
Then, the solver is 
started, postprocessing is performed to extract the diffraction order efficiencies, and results are 
evaluated and saved. 
Figure~\ref{figure_grid} shows a graphical representation of a 3D mesh. 
As numerical settings for the solver, finite elements of fourth-order polynomial degree, automatic settings 
for the transparent boundary conditions, and a rigorous domain-decomposition method (to separate the 
computation of the light field in the multi-stack mirror from the computation in the structured absorber region)
 are chosen. 
Together with a relatively coarse lateral spatial discretization and a fine spatial discretization in $z$-direction
this setting yields discrete problems with between one and two millions of unknowns. 
These problems are solved by direct LU factorization on a computer with extended RAM memory (about 40\,GB)
and several multi-core CPU's (total of 16~cores), 
with typical computation times of about 30\,minutes (per parameter set). 
From comparison to a detailed convergence study on 3D EUV simulations~\cite{Burger2011eom1} 
we estimate that we achieve relative accuracies better than 1\,\%
for all diffraction orders with intensities greater than $10^{-5}$ (relative to the intensity of the incoming beam).

\section{Reconstruction of geometry parameters}
\label{section_results}
\subsection{Simulation results}

\begin{figure}[t]
\begin{center}
  \psfrag{NX}{\sffamily $N_\textrm{x}$}
  \psfrag{NY}{\sffamily $N_\textrm{y}$}
  \includegraphics[width=.4\textwidth]{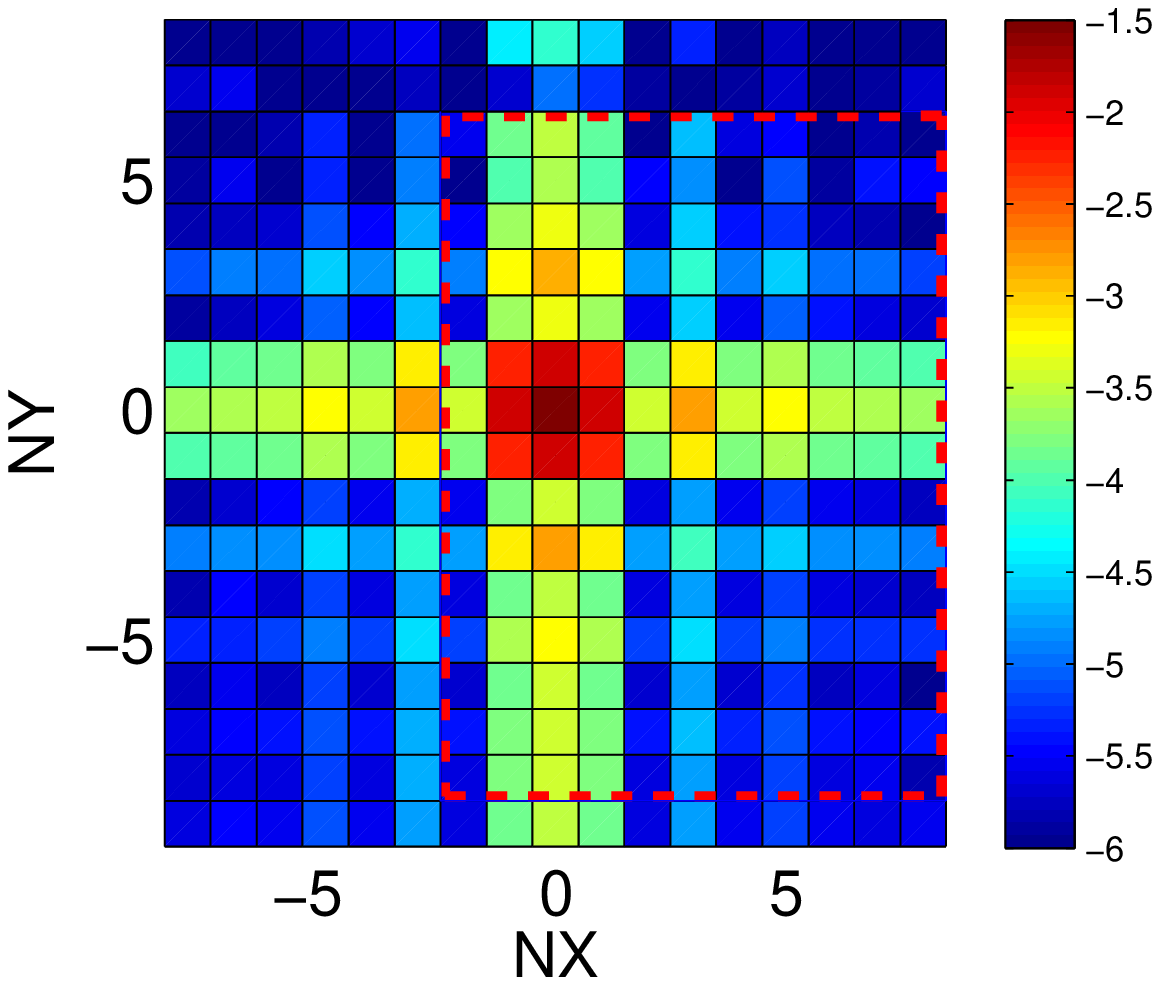}
  \hfill
  \psfrag{Log10(I)}{\sffamily $\log_{10}(I)$}
  \psfrag{Diffraction order}{\sffamily diffraction orders}
  \includegraphics[width=.4\textwidth]{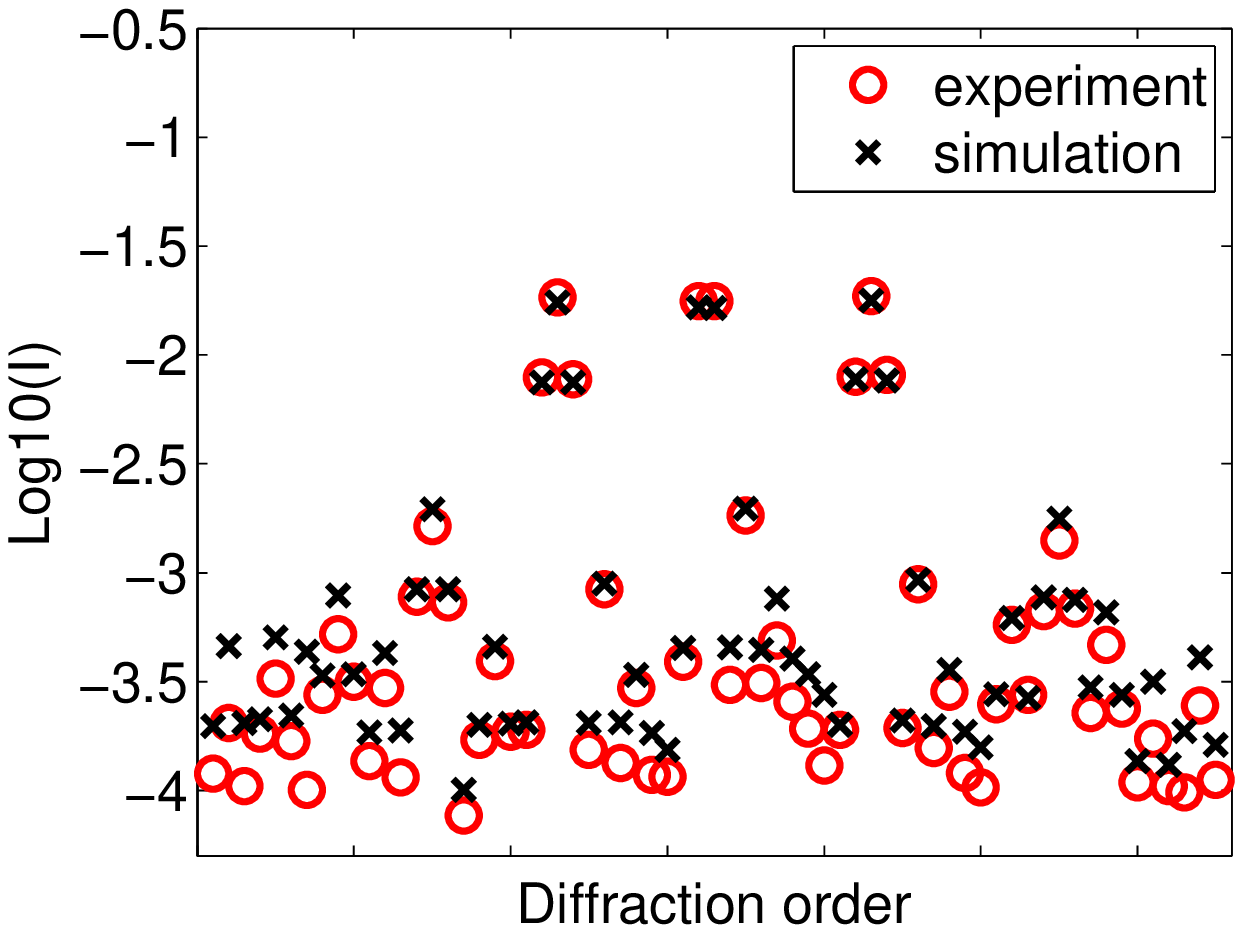}
  \caption{
Scatterogram of  a contact hole array on an EUV mask. 
{\it Left}: Simulated diffraction intensities ($log_{10}(I^\textrm{sim}_{N_\textrm{x}, N_\textrm{y}})$) as function of diffraction index in $x$- and $y$-direction, 
$N_\textrm{x}$, $N_\textrm{y}$ in a pseudo-color representation.
Parameter setting: $CD_\textrm{x} = CD_\textrm{y}=300\,$nm, $\alpha=90\,$deg, $R_{\textrm{xy}}=0\,$nm.
The red, dashed line indicates the range of diffraction orders obtained experimentally. 
{\it Right}: Simulations (x, same data as {\it left}) and corresponding experimental values (o), 
intensities $>10^{-5}$ displayed on a logarithmic scale. 
The 2D array of orders is plotted row by row in $y$ for subsequent
orders in $x$. 
}
\label{simulated_scatterogram}
\end{center}
\end{figure}

The simulated scatterogram of the EUV mask pattern, where all parameters are set to nominal values, is shown in 
Figure~\ref{simulated_scatterogram}. 
The left part of the Figure shows the diffraction intensities on a logarithmic, pseudo-color scale. 
The right part of the Figure displays the same simulation 
results in a 2D graph, together with results from the measurement. 
Only results with measured relative intensities  larger than 
$8\cdot 10^{-5}$ are displayed ({\it cf.,} dashed line in Figure~\ref{figure_experiment_CH}). 

Each simulated diffraction order intensity, $I^\textrm{sim}_{N_\textrm{x}, N_\textrm{y}}$, is attributed a relative deviation:
$$
\delta I^\textrm{sim}_{N_\textrm{x}, N_\textrm{y}}(CD_\textrm{x}, CD_\textrm{y}, \alpha, R_{\textrm{xy}}) = 
\frac{|I^\textrm{sim}_{N_\textrm{x}, N_\textrm{y}}-I^\textrm{exp}_{N_\textrm{x}, N_\textrm{y}}|}
{I^\textrm{exp}_{N_\textrm{x}, N_\textrm{y}}}\,,
$$
where ${I^\textrm{exp}_{N_\textrm{x}, N_\textrm{y}}}$ is the measured value and $N_\textrm{x}$ and $N_\textrm{y}$ denote 
the index of the diffraction order in $x$-, resp.,~$y$-direction. 

A {\it cost function}, is attributed to the simulation parameters by summing over all diffraction orders
$N_\textrm{x}$,  $N_\textrm{y}$, with ${I^\textrm{exp}_{N_\textrm{x}, N_\textrm{y}}}>8*10^{-5}I_0$, where $I_0$ is the intensity 
of the illuminating plane wave:
$$
\delta I^\textrm{sim}(CD_\textrm{x}, CD_\textrm{y}, \alpha, R_{\textrm{xy}}) = 
\sum_{N_\textrm{x}}\sum_{N_\textrm{y}}
\left(
\delta I^\textrm{sim}_{N_\textrm{x}, N_\textrm{y}}(CD_\textrm{x}, CD_\textrm{y}, \alpha, R_{\textrm{xy}})\,/
n_m
\right)^2\, .
$$
Here, $n_m$ denotes the number of diffraction orders taken into account, and the zero$^{\textrm{th}}$ diffraction order
is omitted.

\subsection{Reconstruction results}
\begin{figure}[t]
\begin{center}
  \psfrag{CDx}{\sffamily $CD_\textrm{x}$}
  \psfrag{Rxy}{\sffamily $R_{\textrm{xy}}$}
  \psfrag{cost}{\sffamily $\delta I$}
  \psfrag{Log10(I)}{\sffamily $\log_{10}(I)$}
  \psfrag{Diffraction order}{\sffamily diffraction orders}
  \includegraphics[width=0.325\textwidth]{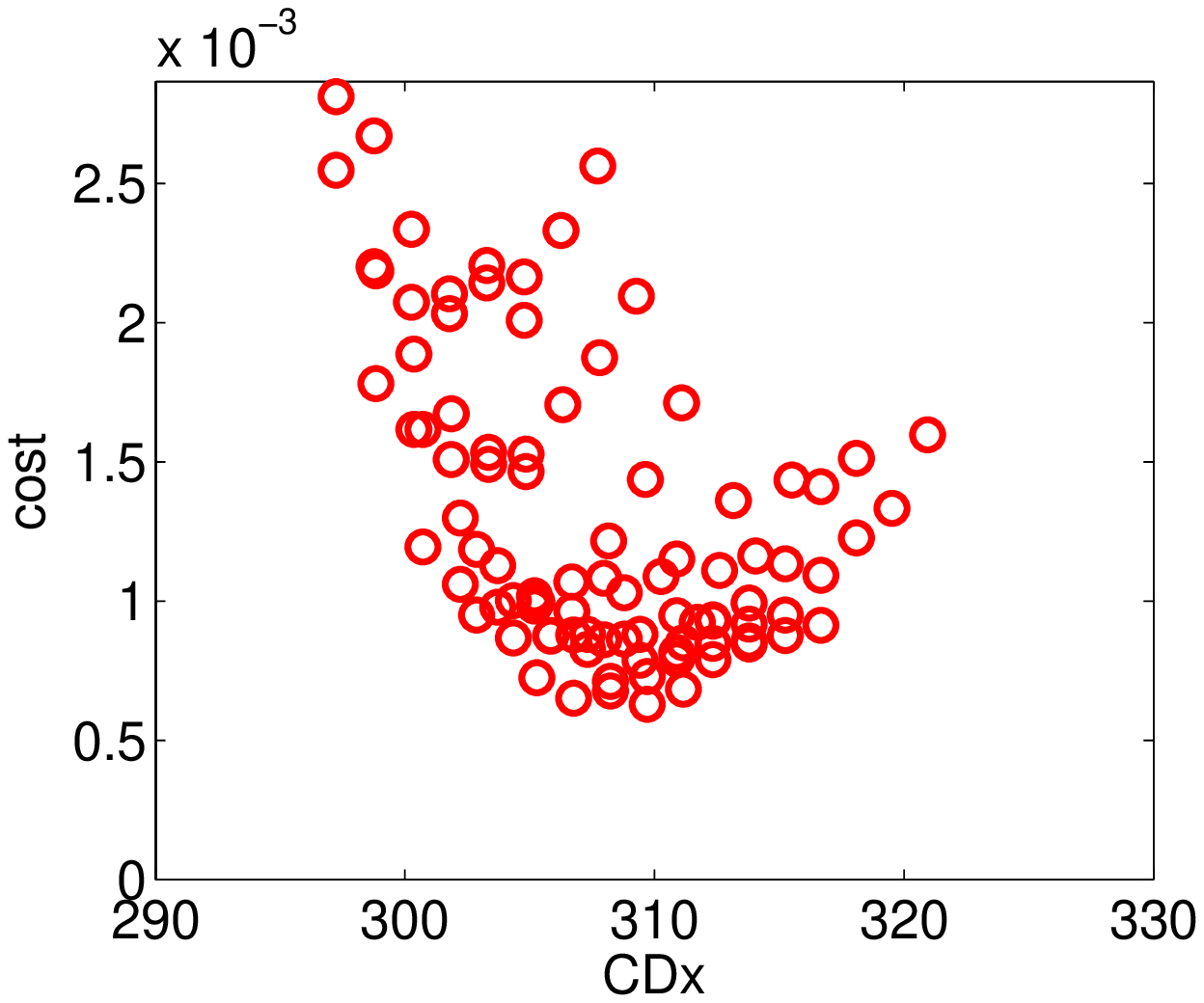}
  \includegraphics[width=0.325\textwidth]{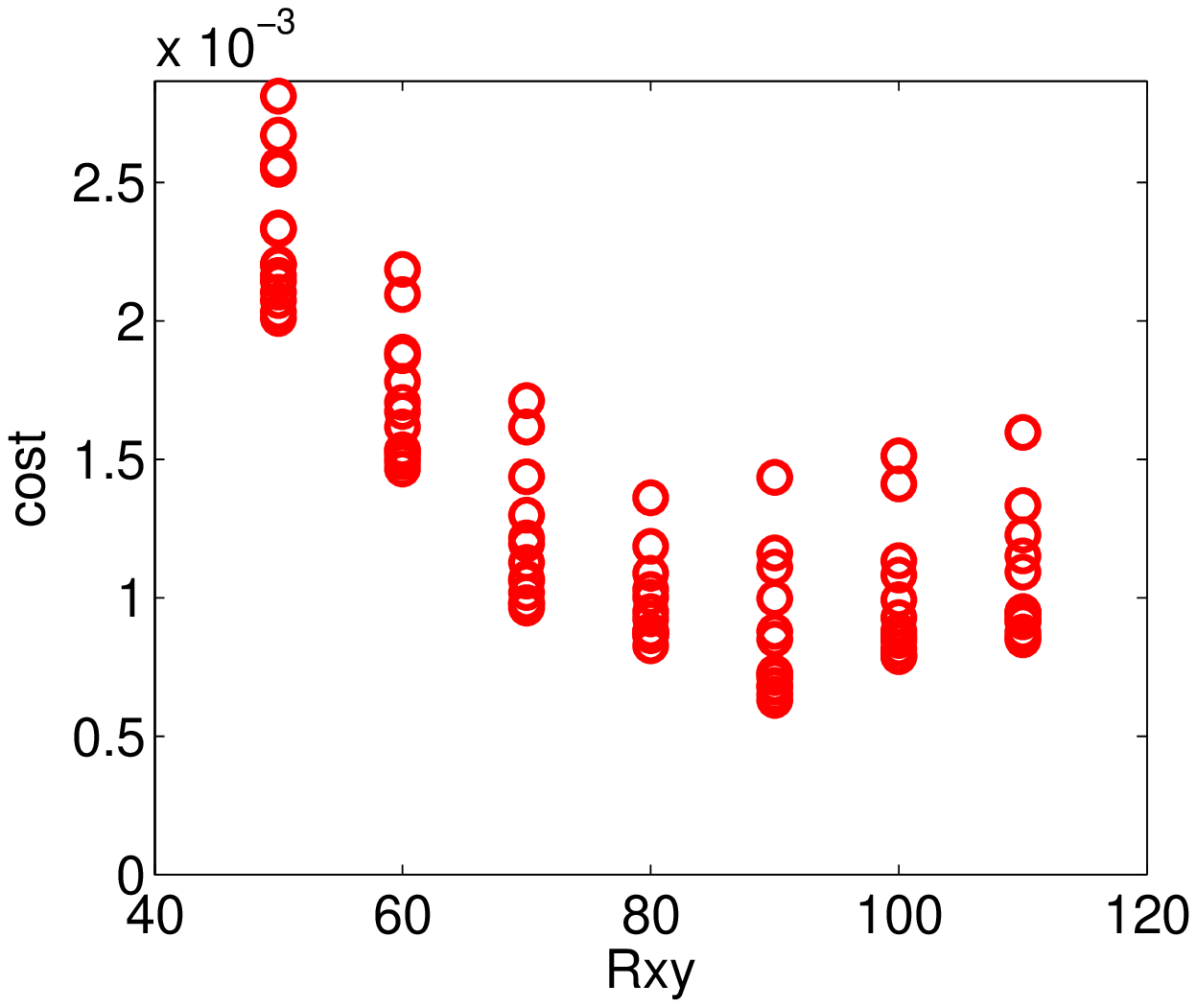}
  \hfill
  \includegraphics[width=0.325\textwidth]{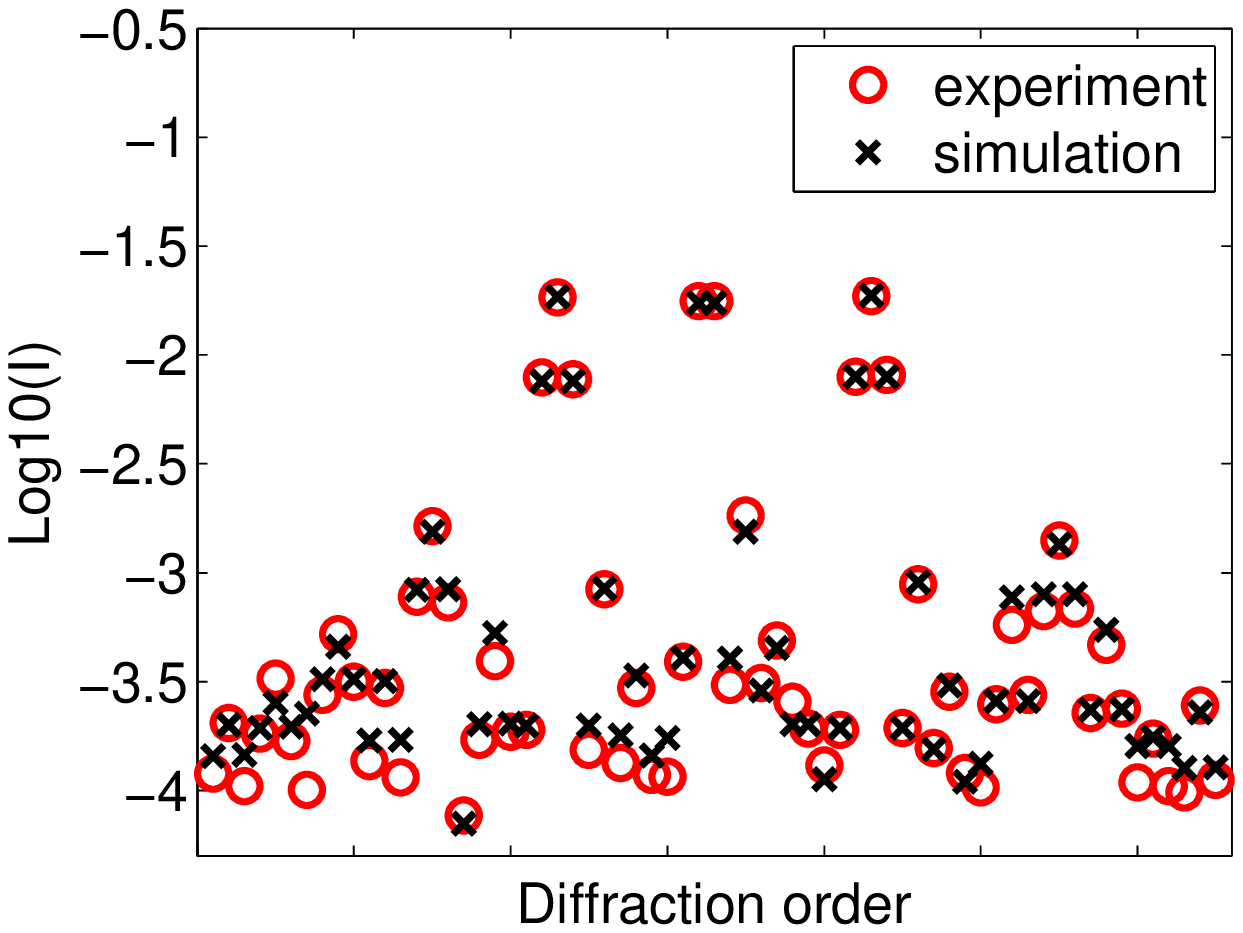}
  \caption{
Reconstruction results. 
{\it Left:} Dependence of cost function on  $CD_\textrm{x}$. 
{\it Center:} Dependence of cost function on  $R_\textrm{xy}$. 
{\it Right:} 
Simulated diffraction intensities (x) for parameter setting with 
highest correlation to experimental data (o).
}
\label{optimization_figure}
\end{center}
\end{figure}

We have performed a series of simulations with varied parameter sets 
($CD_\textrm{x}$, $CD_\textrm{y}$, $\alpha$, $R_{\textrm{xy}}$),
recorded the corresponding diffraction patterns and computed the corresponding cost functions / deviations 
$\delta I^\textrm{sim}$. 
Figure~\ref{optimization_figure} ({\it left, center}) shows how the cost function varies with $CD_\textrm{x}$ and $R_{\textrm{xy}}$, 
in the investigated parameter ranges. 
We find the lowest cost function, i.e., the best correlation between experiment and simulation for the parameter set
 $CD_\textrm{x}=310\,$nm, $CD_\textrm{y}=307\,$nm, $\alpha=88\,$deg, $R_{\textrm{xy}}=90\,$nm. 
The measured and simulated diffraction spectra for this parameter set are displayed in Figure~\ref{optimization_figure} ({\it right}). 
The agreement of most diffraction orders intensities is very good. 
We expect that the parameter values obtained with this scatterometric method are close to the {\it real} parameter 
values. 

However, the fact that the differences between measurement and simulation are larger than both, expected experimental measurement 
uncertainty and numerical errors, suggests the following advancements: \\
(i) More free parameters to be considered in the model (e.g., layer thicknesses $h_\textrm{ARC}, h_\textrm{a}, h_\textrm{b}$). \\
(ii) Larger parameter ranges to be scanned to avoid possibilities of optimizing to a local minimum. \\
(iii) A more refined geometrical model to take into account effects of 
edge roughness~\cite{Kato2010ao,Kato2011eom}, 
surface roughness, corner rounding at the 
bottom corners, etc. \\
(iv) Reconstruction results to be validated by comparison to results obtained with other 
measurement methods, like scanning electron microscopy (SEM) or atomic force microscopy (AFM).

Currently, computation times for the 3D simulations on large computational domains limit the applicability
when many free parameters are taken into account (unless large computational power is available).  
For real-time reconstruction of using the same methods we plan to 
apply the reduced-basis method (RBM) which allows to decrease computation times for rigorous simulations 
of parameterized simulation setups by several orders of magnitude~\cite{Pomplun2010bacus,Kleemann2011eom3}.

\section{Conclusion}
Scatterometric measurements of 2D periodic patterns on EUV masks have been performed using the 
X-ray radiometry beamline of an electron storage ring. 
Rigorous simulations of the measurements have been performed using a finite-element based 
Maxwell solver. 
Very good agreement between experimental results and simulation results has been achieved. 
Parameter reconstruction has been demonstrated. 

For the future we plan to perform measurements of further and more complex patterns, comparison to 
AFM and SEM results, and application of reduced basis simulation methods.

\section*{Acknowledgments}
The test mask was provided by AMTC Dresden within the BMBF project CDuR32. 
We also thank our colleagues Martin Biel, Christian Buchholz, Annett Kampe, 
Jana Puls and Christian Stadelhoff from the EUV beamline for performing the measurements. 
The authors would like to acknowledge the support of
European Regional Development Fund (EFRE) / Investitionsbank Berlin (IBB) through contracts 
ProFIT 10144554 and 10144555.

\bibliography{/home/numerik/bzfburge/texte/biblios/phcbibli,/home/numerik/bzfburge/texte/biblios/group,/home/numerik/bzfburge/texte/biblios/lithography}
\bibliographystyle{spiebib}

\end{document}